\documentclass{article}

\usepackage[final]{ml4mols_2021}

\usepackage[utf8]{inputenc} 
\usepackage[T1]{fontenc}    
\usepackage{natbib}
\usepackage{hyperref}       
\usepackage{url}            
\usepackage{booktabs}       
\usepackage{amsfonts}       
\usepackage{nicefrac}       
\usepackage{microtype}      
\usepackage{graphicx}
\usepackage{algorithm}
\usepackage{algpseudocode}
\graphicspath{ {./figs/} }

\title{FastFlows: Flow-Based Models for Molecular Graph Generation}

\author{%
  Nathan C. Frey \\
  MIT \\
  \texttt{ncfrey@mit.edu} \\
  \AND
Vijay Gadepally \\
MIT \\
\texttt{vijayg@mit.edu} \\
  \And
  Bharath Ramsundar \\
  Deep Forest Sciences \\
  \texttt{bharath@deepforestsci.com} \\
}

\begin{document}

\maketitle

\begin{abstract}
  We propose a framework using normalizing-flow based models, SELF-Referencing Embedded Strings, and multi-objective optimization that efficiently generates small molecules. With an initial training set of only 100 small molecules, FastFlows generates thousands of chemically valid molecules in seconds. Because of the efficient sampling, substructure filters can be applied as desired to eliminate compounds with unreasonable moieties. Using easily computable and learned metrics for druglikeness, synthetic accessibility, and synthetic complexity, we perform a multi-objective optimization to demonstrate how FastFlows functions in a high-throughput virtual screening context. Our model is significantly simpler and easier to train than autoregressive molecular generative models, and enables fast generation and identification of druglike, synthesizable molecules.  
\end{abstract}

\section{Introduction}
The goal of generative modeling of small molecules is to discover structurally novel molecules with optimal physicochemical properties. Prior work using variational autoencoders (VAEs) [\citenum{Gomez-Bombarelli2018}], generative adversarial networks (GANs) [\citenum{DeCao2018}], and reinforcement learning [\citenum{You2018}] has shown the promise of generative modeling in the chemical sciences. Normalizing flows (NFs) [\citenum{Papamakarios2019}] have emerged as a promising model architecture for chemical space distribution learning and molecular graph generation [\citenum{Zang2020, Shi2020, Madhawa2019, Honda2019}]. Unlike the generative models discussed above, NFs do not rely on a compressed latent space representation for generative modeling. Instead, an NF learns an invertible mapping between a simple base distribution and a target distribution.

Previous work applying NFs to molecule generation [\citenum{Zang2020}] has shown that NFs with post-hoc corrections to enforce chemical validity achieve high validity, novelty, and uniqueness scores on benchmark datasets like QM9 [\citenum{Ramakrishnan2014}] and ZINC250K [\citenum{Sterling2015}]. However, NFs require deep architectures with many bijective transformations to model complex target distributions [\citenum{Dinh2016, Papamakarios2017, kingma2018glow}], which results in a prohibitive computational cost to training flows [\citenum{bondtaylor2021deep}]. Some applications of generative modeling may call for learning immense, heterogeneous regions of chemical space, which deep DGMs seem well-suited for. It remains to be seen whether DGMs have any utility in exploring more compact, well-defined chemical spaces, where simpler methods like genetic algorithms [\citenum{nigam2021}] and particle swarm optimization [\citenum{winter2019}] are easier to use and just as, if not more, performative.

In this paper, we present FastFlows, a normalizing flow-based approach for fast and efficient molecular graph sampling with DGMs. Through careful choice of the underlying flow architecture, FastFlows avoids the common difficulties and instabilities of training other generative models like GANs and VAEs. Using the 100\% robust SELF-referencing Embedded Strings (SELFIES) representation [\citenum{Krenn2020}] ensures that all generated samples are chemically valid, so we can train FastFlows in low-data limits (10s or 100s of training datapoints) to "memorize" a target distribution, while still generating novel, unique, and valid samples. Because FastFlows uses a simple molecular graph representation (SELFIES) and does not use autoregressivity, tens of thousands of molecules can be generated each second, avoiding the prohibitive time complexity of sampling from autoregressive DGMs. FastFlows uses substructure filters to discard unreasonable molecules, and easily computed metrics for druglikeness and synthetic accessibility/complexity are used to identify Pareto optimal molecules from thousands of generated candidates. This work presents an effective scheme for practical, robust molecular generation with DGMs, and a path towards discovering novel druglike compounds.

\section{FastFlows: Efficient Flow-Based Generative Models}
\begin{figure}[htbp]
  \centering
  \includegraphics[width=0.6\textwidth]{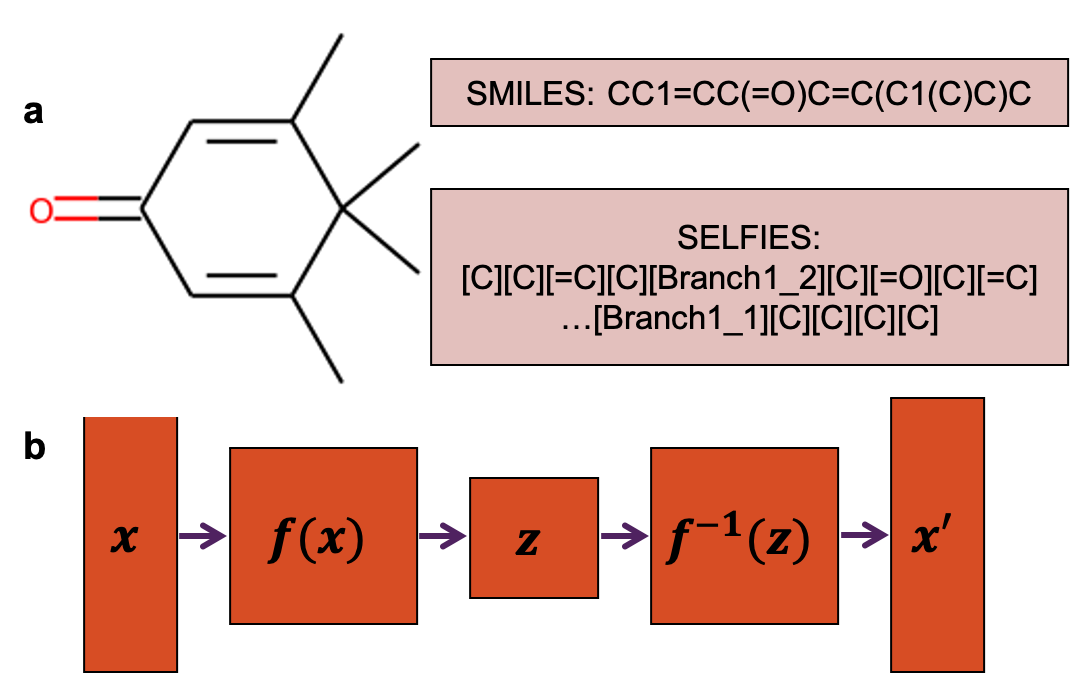}
  \caption{\textbf{a} Representative molecule encoded as SMILES and SELFIES strings. \textbf{b} Normalizing flow architecture that maps an input distribution to a target distribution through invertible transformations.
}
  \label{fig:selfies}
\end{figure}

To circumvent a common problem with molecular generative models - invalid outputs due to chemical rules not being encoded in the model architecture - we first encode molecules as SELFIES strings (Figure \ref{fig:selfies}a). The SELFIES grammar and bond constraints enforce chemical valency rules, guaranteeing that generated SELFIES are syntactically and semantically valid, without requiring post-hoc corrections or complex model architectures that are difficult to train. In a normalizing flow (NF) (Figure \ref{fig:selfies}b), vectors from a simple base distribution like a multi-dimensional Gaussian are passed through the flow, and a sequence of invertible transformations maps the vector to a sample from the target distribution. 

We use an NF to model target distributions of molecules. Because NFs are composed of bijective transformations, they provide an easily interpretable one-to-one mapping between inputs and outputs, without lossy compression to a latent space representation. NFs offer both generative sampling and exact likelihood calculation, unlike VAEs which provide only a lower-bound on log-likelihood and GANs, which do not provide likelihood estimation. Here, the NF provides a simple and straightforward means of generating new molecules.

\begin{figure}[htbp]
  \includegraphics[width=0.6\textwidth]{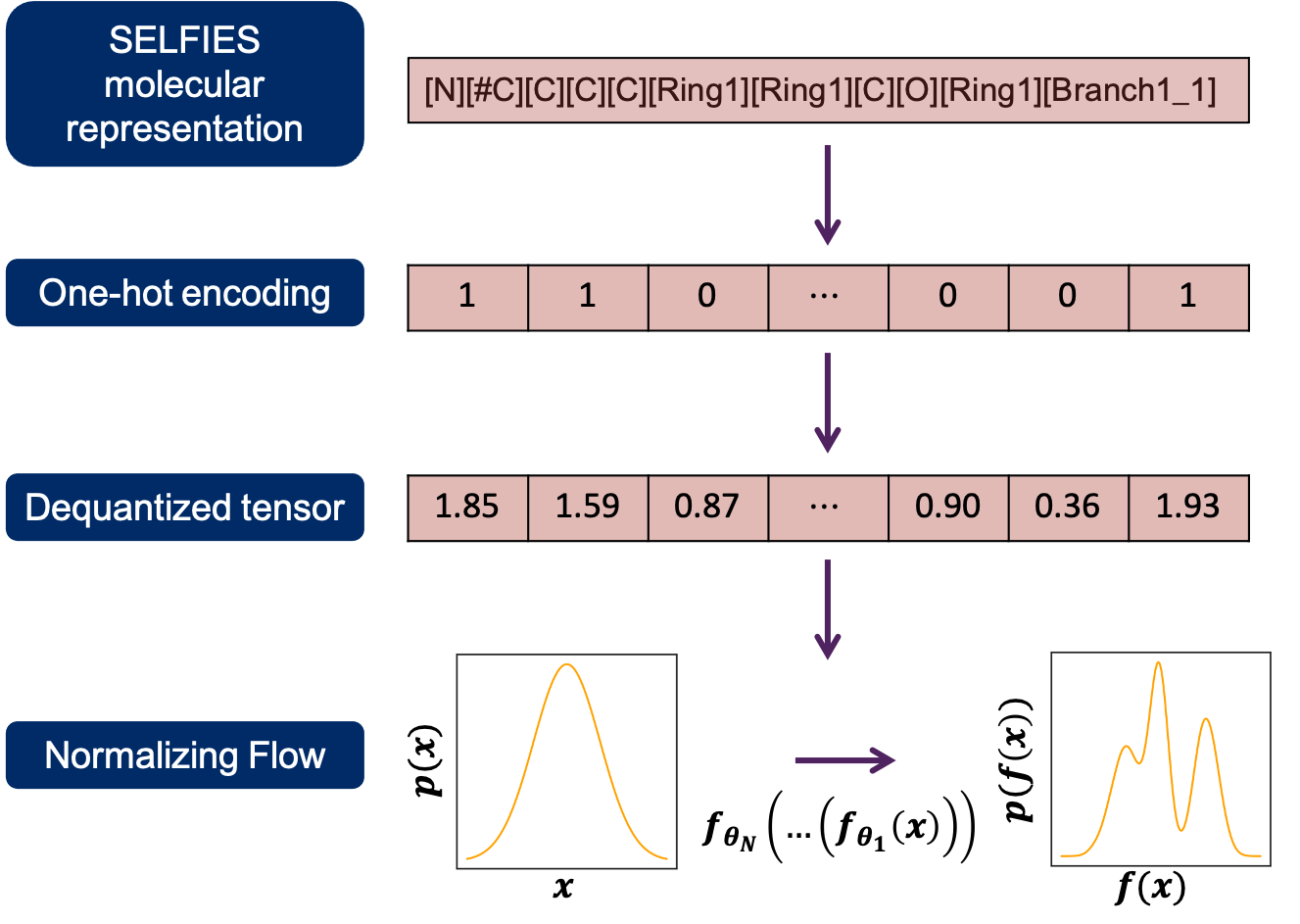}
  \centering
  \caption{Dequantized one-hot encodings of SELFIES representations are inputs to the normalizing flow. The normalizing flow maps a simple base distribution to a complex target distribution.}
  \label{fig:training}
\end{figure}

A schematic of the data pre-processing steps and mapping between the base and target distributions is shown in Figure \ref{fig:training}. SELFIES strings are one-hot encoded and dequantized [\citenum{Dinh2016}] by adding random noise from the interval $[0, 0.95)$ to each element. The original inputs can be recovered by applying a floor function, and the continuous dequantized inputs are used to train the model. The NF uses real-valued non-volume preserving (Real NVP) transformations [\citenum{Dinh2016}] with 32 layers and 8 residual blocks per layer with 16 hidden feature maps, and checkerboard masking. 

Real NVP has the advantages of efficient and fast training and sampling, unlike autoregressive models, because sampling is parallelized over input dimensions [\citenum{Dinh2016}]. Sampling efficiency is a key factor in enabling FastFlows, although Real NVP is less expressive and requires a deeper network architecture than more recently developed autoregressive and residual flows [\citenum{Papamakarios2017, chen2020residual}]. The base distribution is a multi-dimensional standard normal. When training FastFlows, a small batch size of 4 is used, the learning rate is set to 1e-3, and the dataset is restricted to 100 samples from the target distribution. 

\begin{figure}[htbp]
  \includegraphics[width=0.5\textwidth]{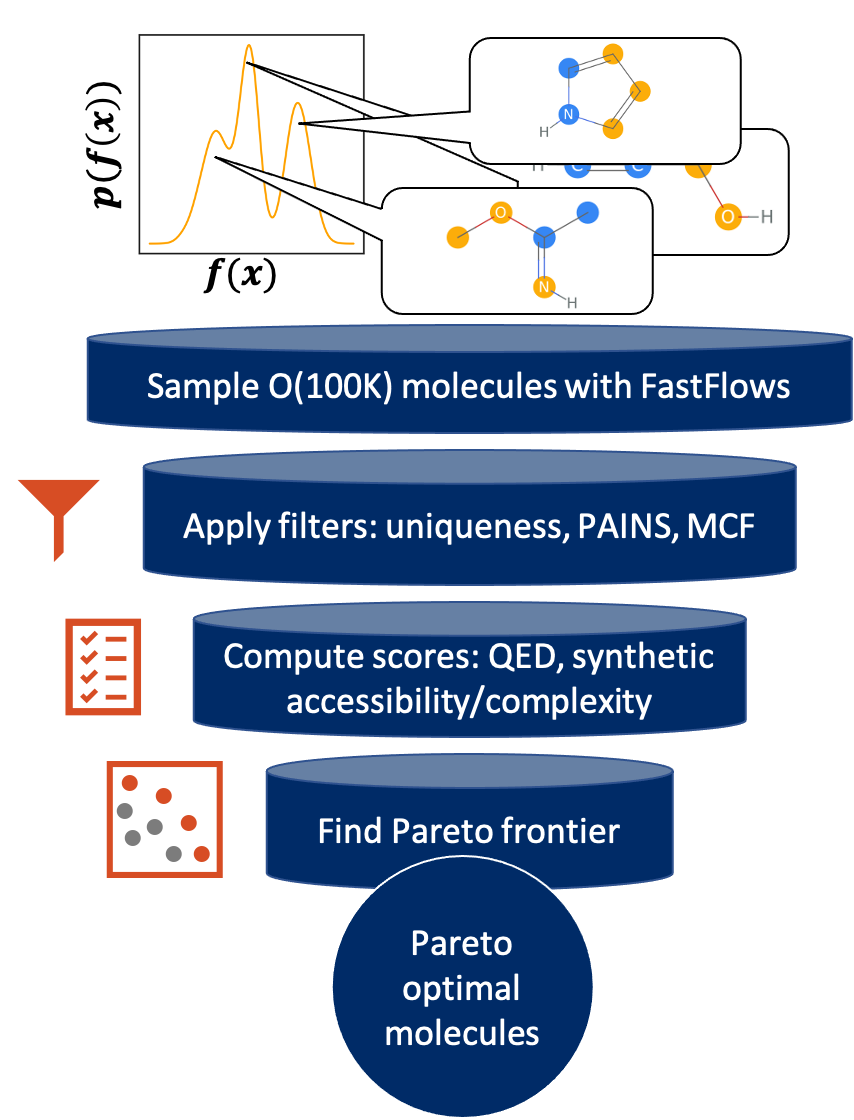}
  \centering
  \caption{FastFlows workflow diagram. 100K molecules are sampled from a trained FastFlow in 4.2 seconds. Substructure filters are applied as needed, and objective targets are calculated for each sample. Pareto optimal samples are identified according to trade-offs between druglikeness and synthesizability.
}
  \label{fig:workflow}
\end{figure}

The molecular graph generation workflow is shown schematically in Figure \ref{fig:workflow}. The generative model is trained on an initial subset of training data. Due to the parallelized, efficient sampling of Real NVP, we generate samples and transform them back into SMILES strings at a rate of 23,981 molecules/second on a single NVIDIA Tesla V100 GPU. Samples are filtered for uniqueness, pan-assay interfering compounds [\citenum{baell2010}], and unstable or reactive moieties [\citenum{polykovskiy2020molecular}]. Again, because of the cheap computational and time costs of molecular generation, we can generate chemically valid samples from the target distribution and filter out undesirable compounds as needed. Using a more sophisticated model architecture or larger training set may increase the diversity and unique percentage of generated samples, but FastFlows is designed for the practical purpose of rapid sampling that enables high-throughput virtual screening of compact target distributions, rather than generalized distribution learning. 

For simplicity, we score generated molecules with quantitative estimate of druglikeness (QED) [\citenum{Bickerton2012}], synthetic accessibility (SA) [\citenum{Ertl2009}], and a synthetic complexity score (SCScore) [\citenum{coley2018}] learned from a corpus of chemical reactions. The scored molecules are used to find the Pareto frontier of optimal candidates, using the \emph{PyePAL} [\citenum{Jablonka2021}] package. The algorithm for finding the Pareto frontier is given in Algorithm \ref{alg:alg1}. A molecule is Pareto dominated if there is another molecule that is at least as performative along every metric, and more performative for at least one metric. Pareto optimal molecules define the Pareto frontier, where target metrics cannot be improved without degrading others. 

\begin{algorithm}
\caption{Pareto frontier algorithm.}\label{alg:alg1}
\begin{algorithmic}
\State \textbf{Input:} generated molecule score values
\For{scores \emph{s} in score list}
    \If{any score > s}  \Comment{Find if point is dominated}
     \State add s to dominated list
    \EndIf
\EndFor
\State \textbf{Return:} dominated list
\end{algorithmic}
\end{algorithm}

For example, we might intuitively expect that synthetically complex molecules are less synthetically accessible, which introduces a trade-off between the two properties. The Pareto frontier solves the multi-objective optimization problem of identifying generated molecules that are maximally druglike, synthesizable, and complex (more like a reaction product than a reactant). We also note that QED and synthesizability metrics are correlated with model quality [\citenum{Skinnider2021}], whereas many other commonly used generative model metrics can be trivially satisfied by a judicious choice of molecular representation [\citenum{nigam2021}], or are not correlated with model quality [\citenum{Skinnider2021}].

\section{Experiments}

\begin{figure}[htbp]
  \includegraphics[width=0.8\textwidth]{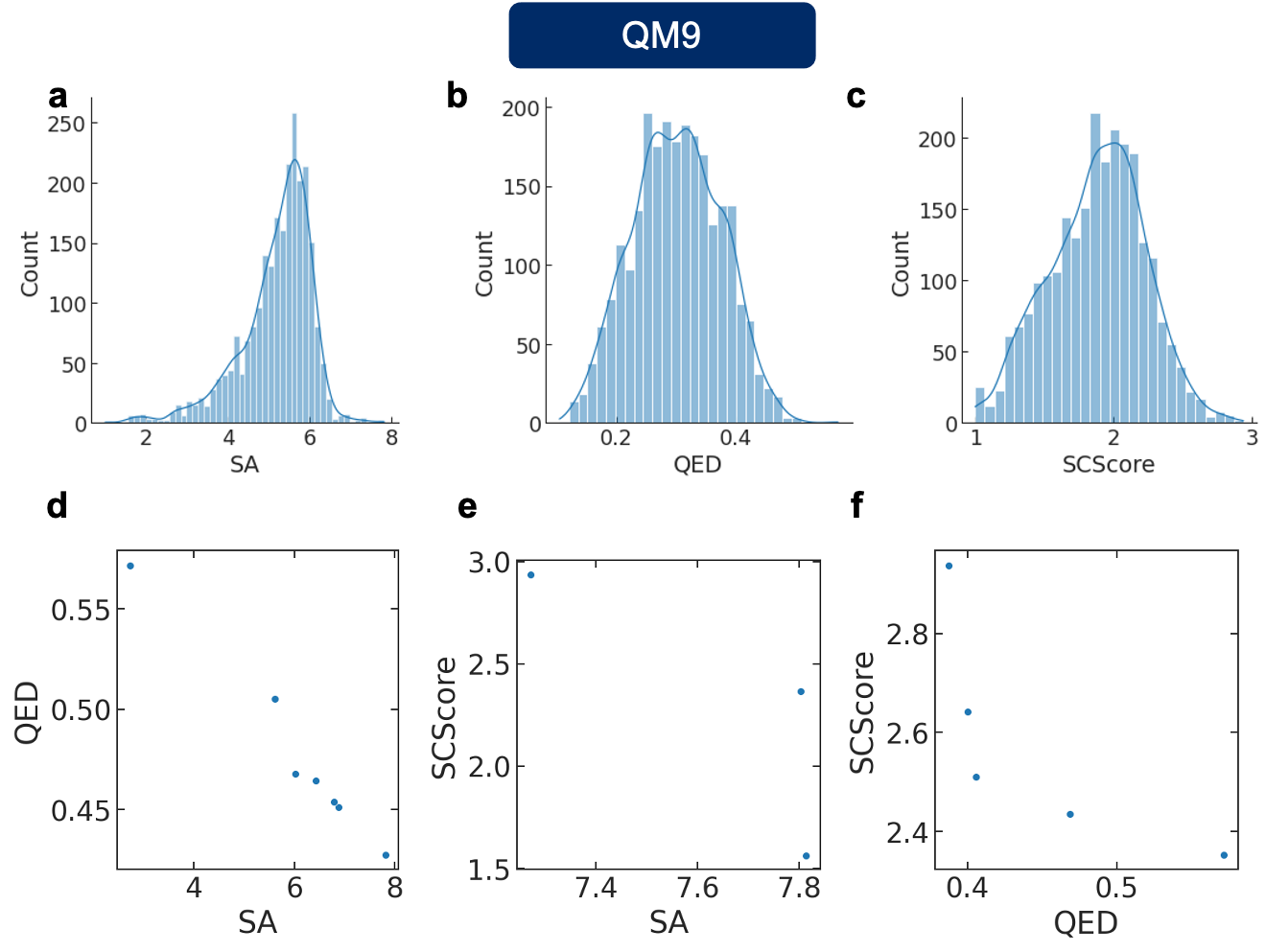}
  \centering
  \caption{Distributions of \textbf{a} synthetic accessibility, \textbf{b} quantitative estimate of druglikeness, and \textbf{c} synthetic complexity score for generated samples. Pareto frontiers (\textbf{d}-\textbf{f}) for generated samples.}
  \label{fig:qm9_pareto}
\end{figure}

\paragraph{QM9}
Our goal is to provide a framework for solving the multi-objective optimization problem of finding druglike, synthetically accessible, and synthetically complex molecules. While optimizing for single objectives may lead to molecules that are too simple (synthetic accessibility), too similar to already-known drugs (QED), or too complex (SCScore), simultaneously optimizing for all three metrics leads to a richer exploration of the generated samples. Figure \ref{fig:qm9_pareto}\textbf{a}-\textbf{c} shows the distributions of SA, QED, and SCScore for over 2,000 generated molecules after applying filters.

Two-dimensional Pareto frontiers are shown in Figure \ref{fig:qm9_pareto}\textbf{d}-\textbf{f}, comparing trade-offs between each combination of metrics. In Figure \ref{fig:qm9_pareto}e, we observe that calculating the Pareto frontier with respect to SCScore and SA reduces the candidate molecule space from over 2,000 to only three molecules. Each frontier provides a short list of Pareto optimal candidates for further investigation, via more computationally demanding techniques like docking and molecular dynamics simulations.


\paragraph{ChEMBL}
We repeat the above FastFlow experiment trained on 100 compounds from ChEMBL [\citenum{mendez2018}]. ChEMBL molecules are, on average, much larger than those from QM9, which impacts training and generation. After training on ChEMBL molecules, FastFlows generates 481 molecules per second (versus nearly 24,000 molecules per second when trained on QM9). Still, the extremely fast sampling means that sampling molecules is memory-limited rather than time-limited. 

While the FastFlows approach extends to more relevant target distributions for drug discovery like ChEMBL, the increased dimensionality of interesting chemical spaces presents a difficulty for flow-based models. For 100 random samples from ChEMBL, the largest SELFIES string is 552 characters and the SELFIES vocabulary includes 33 characters, yielding a target distribution with dimensionality of 18216. Mapping the base 18216-dimensional standard normal distribution to the target with an NF requires many successive bijective transformations, and again highlights the need for more expressive flows that preserve fast sampling and training.

\section{Discussion}

In this work, we presented FastFlows, a framework for generative modeling of small molecules using normalizing flows. We showed that lightweight normalizing flows trained in the low-data limit on SELFIES representations can be used as computationally cheap, efficient deep generative models for sampling chemical matter from target distributions. Fast training and the ability to sample more than 20,000 molecules / second enable coupling of the generative model to downstream filtering and multi-objective optimization to identify chemically valid, unique, synthetically accessible, and complex druglike molecules. We demonstrated this workflow on target distributions from the QM9 and ChEMBL datasets. Importantly, the code used for deep generative modeling in this work is available in the \emph{DeepChem} package [\citenum{Ramsundar2019}] implemented with TensorFlowProbability [\citenum{dillon2017tensorflow}], and in the \emph{nflows} PyTorch library [\citenum{nflows}] and the SELFIES library [\citenum{Krenn2020}] at \url{https://github.com/aspuru-guzik-group/selfies}. We identified challenges with using normalizing flows for distribution learning, namely, the need for deep networks with many bijective transformations to adequately learn a mapping to the target distribution. For this reason, it is prohibitive from a computational cost standpoint for current flow-based models to be competitive with other architectures on distribution learning tasks [\citenum{bondtaylor2021deep}]. 

\vspace{-3mm}

\setcitestyle{numbers}
\bibliographystyle{unsrtnat} 
\bibliography{bib}

\begin{thebibliography}{29}
\providecommand{\natexlab}[1]{#1}
\providecommand{\url}[1]{\texttt{#1}}
\expandafter\ifx\csname urlstyle\endcsname\relax
  \providecommand{\doi}[1]{doi: #1}\else
  \providecommand{\doi}{doi: \begingroup \urlstyle{rm}\Url}\fi

\bibitem[G{\'{o}}mez-Bombarelli et~al.(2018)G{\'{o}}mez-Bombarelli, Wei,
  Duvenaud, Hern{\'{a}}ndez-Lobato, S{\'{a}}nchez-Lengeling, Sheberla,
  Aguilera-Iparraguirre, Hirzel, Adams, and Aspuru-Guzik]{Gomez-Bombarelli2018}
Rafael G{\'{o}}mez-Bombarelli, Jennifer~N. Wei, David Duvenaud,
  Jos{\'{e}}~Miguel Hern{\'{a}}ndez-Lobato, Benjam{\'{i}}n
  S{\'{a}}nchez-Lengeling, Dennis Sheberla, Jorge Aguilera-Iparraguirre,
  Timothy~D. Hirzel, Ryan~P. Adams, and Al{\'{a}}n Aspuru-Guzik.
\newblock {Automatic Chemical Design Using a Data-Driven Continuous
  Representation of Molecules}.
\newblock \emph{ACS Central Science}, 4\penalty0 (2):\penalty0 268--276, feb
  2018.
\newblock ISSN 23747951.
\newblock \doi{10.1021/acscentsci.7b00572}.

\bibitem[{De Cao} and Kipf(2018)]{DeCao2018}
Nicola {De Cao} and Thomas Kipf.
\newblock {MolGAN: An implicit generative model for small molecular graphs}.
\newblock 2018.
\newblock URL \url{http://arxiv.org/abs/1805.11973}.

\bibitem[You et~al.(2018)You, Liu, Ying, Pande, and Leskovec]{You2018}
Jiaxuan You, Bowen Liu, Rex Ying, Vijay Pande, and Jure Leskovec.
\newblock {Graph convolutional policy network for goal-directed molecular graph
  generation}.
\newblock In \emph{Advances in Neural Information Processing Systems}, volume
  2018-Decem, pages 6410--6421, 2018.

\bibitem[Papamakarios et~al.(2019)Papamakarios, Nalisnick, Rezende, Mohamed,
  and Lakshminarayanan]{Papamakarios2019}
George Papamakarios, Eric Nalisnick, Danilo~Jimenez Rezende, Shakir Mohamed,
  and Balaji Lakshminarayanan.
\newblock {Normalizing Flows for Probabilistic Modeling and Inference}.
\newblock 2019.
\newblock URL \url{http://arxiv.org/abs/1912.02762}.

\bibitem[Zang and Wang(2020)]{Zang2020}
Chengxi Zang and Fei Wang.
\newblock {MoFlow: An Invertible Flow Model for Generating Molecular Graphs}.
\newblock In \emph{Proceedings of the ACM SIGKDD International Conference on
  Knowledge Discovery and Data Mining}, pages 617--626, 2020.
\newblock ISBN 9781450379984.
\newblock \doi{10.1145/3394486.3403104}.

\bibitem[Shi et~al.(2020)Shi, Xu, Zhu, Zhang, Zhang, and Tang]{Shi2020}
Chence Shi, Minkai Xu, Zhaocheng Zhu, Weinan Zhang, Ming Zhang, and Jian Tang.
\newblock {GraphAF: a Flow-based Autoregressive Model for Molecular Graph
  Generation}.
\newblock 2020.
\newblock URL \url{https://github.com/DeepGraphLearning/GraphAF
  http://arxiv.org/abs/2001.09382}.

\bibitem[Madhawa et~al.(2019)Madhawa, Ishiguro, Nakago, and Abe]{Madhawa2019}
Kaushalya Madhawa, Katushiko Ishiguro, Kosuke Nakago, and Motoki Abe.
\newblock {GraphNVP: An Invertible Flow Model for Generating Molecular Graphs}.
\newblock may 2019.
\newblock URL \url{http://arxiv.org/abs/1905.11600}.

\bibitem[Honda et~al.(2019)Honda, Akita, Ishiguro, Nakanishi, and
  Oono]{Honda2019}
Shion Honda, Hirotaka Akita, Katsuhiko Ishiguro, Toshiki Nakanishi, and Kenta
  Oono.
\newblock {Graph Residual Flow for Molecular Graph Generation}.
\newblock 2019.
\newblock URL \url{http://arxiv.org/abs/1909.13521}.

\bibitem[Ramakrishnan et~al.(2014)Ramakrishnan, Dral, Rupp, and {Von
  Lilienfeld}]{Ramakrishnan2014}
Raghunathan Ramakrishnan, Pavlo~O. Dral, Matthias Rupp, and O.~Anatole {Von
  Lilienfeld}.
\newblock {Quantum chemistry structures and properties of 134 kilo molecules}.
\newblock \emph{Scientific Data}, 1\penalty0 (1):\penalty0 1--7, aug 2014.
\newblock ISSN 20524463.
\newblock \doi{10.1038/sdata.2014.22}.
\newblock URL \url{www.nature.com/sdata/}.

\bibitem[Sterling and Irwin(2015)]{Sterling2015}
Teague Sterling and John~J. Irwin.
\newblock {ZINC 15 - Ligand Discovery for Everyone}.
\newblock \emph{Journal of Chemical Information and Modeling}, 55\penalty0
  (11):\penalty0 2324--2337, nov 2015.
\newblock ISSN 15205142.
\newblock \doi{10.1021/acs.jcim.5b00559}.
\newblock URL \url{https://clinicaltrials.gov}.

\bibitem[Dinh et~al.(2016)Dinh, Sohl-Dickstein, and Bengio]{Dinh2016}
Laurent Dinh, Jascha Sohl-Dickstein, and Samy Bengio.
\newblock {Density estimation using Real NVP}.
\newblock \emph{5th International Conference on Learning Representations, ICLR
  2017 - Conference Track Proceedings}, may 2016.
\newblock URL \url{http://arxiv.org/abs/1605.08803}.

\bibitem[Papamakarios et~al.(2017)Papamakarios, Pavlakou, and
  Murray]{Papamakarios2017}
George Papamakarios, Theo Pavlakou, and Iain Murray.
\newblock {Masked Autoregressive Flow for Density Estimation}.
\newblock \emph{Advances in Neural Information Processing Systems},
  2017-December:\penalty0 2339--2348, may 2017.
\newblock URL \url{http://arxiv.org/abs/1705.07057}.

\bibitem[Kingma and Dhariwal(2018)]{kingma2018glow}
Diederik~P. Kingma and Prafulla Dhariwal.
\newblock Glow: Generative flow with invertible 1x1 convolutions, 2018.

\bibitem[Bond-Taylor et~al.(2021)Bond-Taylor, Leach, Long, and
  Willcocks]{bondtaylor2021deep}
Sam Bond-Taylor, Adam Leach, Yang Long, and Chris~G. Willcocks.
\newblock Deep generative modelling: A comparative review of vaes, gans,
  normalizing flows, energy-based and autoregressive models, 2021.

\bibitem[Nigam et~al.(2021)Nigam, Pollice, Krenn, Gomes, and
  Aspuru-Guzik]{nigam2021}
AkshatKumar Nigam, Robert Pollice, Mario Krenn, Gabriel dos~Passos Gomes, and
  Alán Aspuru-Guzik.
\newblock Beyond generative models: superfast traversal{,} optimization{,}
  novelty{,} exploration and discovery (stoned) algorithm for molecules using
  selfies.
\newblock \emph{Chem. Sci.}, 12:\penalty0 7079--7090, 2021.
\newblock \doi{10.1039/D1SC00231G}.
\newblock URL \url{http://dx.doi.org/10.1039/D1SC00231G}.

\bibitem[Winter et~al.(2019)Winter, Montanari, Steffen, Briem, Noé, and
  Clevert]{winter2019}
Robin Winter, Floriane Montanari, Andreas Steffen, Hans Briem, Frank Noé, and
  Djork-Arné Clevert.
\newblock Efficient multi-objective molecular optimization in a continuous
  latent space.
\newblock \emph{Chem. Sci.}, 10:\penalty0 8016--8024, 2019.
\newblock \doi{10.1039/C9SC01928F}.
\newblock URL \url{http://dx.doi.org/10.1039/C9SC01928F}.

\bibitem[Krenn et~al.(2020)Krenn, Hase, Nigam, Friederich, and
  Aspuru-Guzik]{Krenn2020}
Mario Krenn, Florian Hase, Akshatkumar Nigam, Pascal Friederich, and Alan
  Aspuru-Guzik.
\newblock {Self-Referencing Embedded Strings (SELFIES): A 100{\%} robust
  molecular string representation}.
\newblock \emph{Machine Learning: Science and Technology}, 2020.
\newblock \doi{10.1088/2632-2153/aba947}.

\bibitem[Chen et~al.(2020)Chen, Behrmann, Duvenaud, and
  Jacobsen]{chen2020residual}
Ricky T.~Q. Chen, Jens Behrmann, David Duvenaud, and Jörn-Henrik Jacobsen.
\newblock Residual flows for invertible generative modeling, 2020.

\bibitem[Baell and Holloway(2010)]{baell2010}
Jonathan~B. Baell and Georgina~A. Holloway.
\newblock New substructure filters for removal of pan assay interference
  compounds (pains) from screening libraries and for their exclusion in
  bioassays.
\newblock \emph{Journal of Medicinal Chemistry}, 53\penalty0 (7):\penalty0
  2719--2740, 2010.
\newblock \doi{10.1021/jm901137j}.
\newblock URL \url{https://doi.org/10.1021/jm901137j}.
\newblock PMID: 20131845.

\bibitem[Polykovskiy et~al.(2020)Polykovskiy, Zhebrak, Sanchez-Lengeling,
  Golovanov, Tatanov, Belyaev, Kurbanov, Artamonov, Aladinskiy, Veselov,
  Kadurin, Johansson, Chen, Nikolenko, Aspuru-Guzik, and
  Zhavoronkov]{polykovskiy2020molecular}
Daniil Polykovskiy, Alexander Zhebrak, Benjamin Sanchez-Lengeling, Sergey
  Golovanov, Oktai Tatanov, Stanislav Belyaev, Rauf Kurbanov, Aleksey
  Artamonov, Vladimir Aladinskiy, Mark Veselov, Artur Kadurin, Simon Johansson,
  Hongming Chen, Sergey Nikolenko, Alan Aspuru-Guzik, and Alex Zhavoronkov.
\newblock Molecular sets (moses): A benchmarking platform for molecular
  generation models, 2020.

\bibitem[Bickerton et~al.(2012)Bickerton, Paolini, Besnard, Muresan, and
  Hopkins]{Bickerton2012}
G.~Richard Bickerton, Gaia~V. Paolini, J{\'{e}}r{\'{e}}my Besnard, Sorel
  Muresan, and Andrew~L. Hopkins.
\newblock {Quantifying the chemical beauty of drugs}.
\newblock \emph{Nature Chemistry}, 4\penalty0 (2):\penalty0 90--98, feb 2012.
\newblock ISSN 17554330.
\newblock \doi{10.1038/nchem.1243}.
\newblock URL \url{www.nature.com/naturechemistry}.

\bibitem[Ertl and Schuffenhauer(2009)]{Ertl2009}
Peter Ertl and Ansgar Schuffenhauer.
\newblock Estimation of synthetic accessibility score of drug-like molecules
  based on molecular complexity and fragment contributions.
\newblock \emph{Journal of Cheminformatics}, 1, 2009.
\newblock ISSN 17582946.
\newblock \doi{10.1186/1758-2946-1-8}.

\bibitem[Coley et~al.(2018)Coley, Rogers, Green, and Jensen]{coley2018}
Connor~W. Coley, Luke Rogers, William~H. Green, and Klavs~F. Jensen.
\newblock Scscore: Synthetic complexity learned from a reaction corpus.
\newblock \emph{Journal of Chemical Information and Modeling}, 58\penalty0
  (2):\penalty0 252--261, 2018.
\newblock \doi{10.1021/acs.jcim.7b00622}.
\newblock URL \url{https://doi.org/10.1021/acs.jcim.7b00622}.
\newblock PMID: 29309147.

\bibitem[Jablonka et~al.(2021)Jablonka, Jothiappan, Wang, Smit, and
  Yoo]{Jablonka2021}
Kevin~Maik Jablonka, Giriprasad~Melpatti Jothiappan, Shefang Wang, Berend Smit,
  and Brian Yoo.
\newblock Bias free multiobjective active learning for materials design and
  discovery.
\newblock \emph{Nature Communications}, 12:\penalty0 2312, 12 2021.
\newblock ISSN 2041-1723.
\newblock \doi{10.1038/s41467-021-22437-0}.
\newblock URL \url{http://www.nature.com/articles/s41467-021-22437-0}.

\bibitem[Skinnider et~al.(2021)Skinnider, Stacey, Wishart, and
  Foster]{Skinnider2021}
Michael~A. Skinnider, R.~Greg Stacey, David~S. Wishart, and Leonard~J. Foster.
\newblock Chemical language models enable navigation in sparsely populated
  chemical space.
\newblock \emph{Nature Machine Intelligence}, 2021.
\newblock ISSN 25225839.
\newblock \doi{10.1038/s42256-021-00368-1}.

\bibitem[Mendez et~al.(2018)Mendez, Gaulton, Bento, Chambers, De Veij, Félix,
  Magariños, Mosquera, Mutowo, Nowotka, Gordillo-Marañón, Hunter, Junco,
  Mugumbate, Rodriguez-Lopez, Atkinson, Bosc, Radoux, Segura-Cabrera, Hersey,
  and Leach]{mendez2018}
David Mendez, Anna Gaulton, A~Patrícia Bento, Jon Chambers, Marleen De Veij,
  Eloy Félix, María Paula Magariños, Juan F Mosquera, Prudence Mutowo,
  Michał Nowotka, María Gordillo-Marañón, Fiona Hunter, Laura Junco, Grace
  Mugumbate, Milagros Rodriguez-Lopez, Francis Atkinson, Nicolas Bosc, Chris J
  Radoux, Aldo Segura-Cabrera, Anne Hersey, and Andrew R Leach.
\newblock {ChEMBL: towards direct deposition of bioassay data}.
\newblock \emph{Nucleic Acids Research}, 47\penalty0 (D1):\penalty0 D930--D940,
  11 2018.
\newblock ISSN 0305-1048.
\newblock \doi{10.1093/nar/gky1075}.
\newblock URL \url{https://doi.org/10.1093/nar/gky1075}.

\bibitem[Ramsundar et~al.(2019)Ramsundar, Eastman, Walters, and
  Pande]{Ramsundar2019}
Bharath Ramsundar, Peter Eastman, Patrick Walters, and Vijay Pande.
\newblock \emph{{Deep Learning for the Life Sciences: Applying Deep Learning to
  Genomics, Drug Discovery, and More}}.
\newblock 2019.
\newblock ISBN 9788578110796.
\newblock \doi{10.1017/CBO9781107415324.004}.
\newblock URL
  \url{https://www.amazon.com/Deep-Learning-Life-Sciences-Microscopy/dp/1492039837}.

\bibitem[Dillon et~al.(2017)Dillon, Langmore, Tran, Brevdo, Vasudevan, Moore,
  Patton, Alemi, Hoffman, and Saurous]{dillon2017tensorflow}
Joshua~V. Dillon, Ian Langmore, Dustin Tran, Eugene Brevdo, Srinivas Vasudevan,
  Dave Moore, Brian Patton, Alex Alemi, Matt Hoffman, and Rif~A. Saurous.
\newblock Tensorflow distributions, 2017.

\bibitem[Durkan et~al.(2020)Durkan, Bekasov, Murray, and Papamakarios]{nflows}
Conor Durkan, Artur Bekasov, Iain Murray, and George Papamakarios.
\newblock {nflows}: normalizing flows in {PyTorch}, November 2020.
\newblock URL \url{https://doi.org/10.5281/zenodo.4296287}.

\end{thebibliography}

\end{document}